\documentstyle[sprocl,epsf]{article}

\def\Journal#1#2#3#4{{#1} {\bf #2}, #3 (#4)}

\def\NPA{{\em Nucl. Phys.} A}

\def\PLB{{\em Phys. Lett.}  B}
\def\PRL{\em Phys. Rev. Lett.}
\def\PRC{{\em Phys. Rev.} C}
\def\PRD{{\em Phys. Rev.} D}
\def\PR{{\em Phys. Rep.}}
\def\ZPC{{\em Z. Phys.} C}
\def\EPJC{{\em Eur. Phys. J.} C}
\def\JPG{{\em J. Phys.} G}
\def\HIP{{\em Heavy Ion Phys.}}

\def\al{\alpha}

\def\be{\begin{equation}}
\def\ee{\end{equation}}
\def\bea{\begin{eqnarray}}
\def\eea{\end{eqnarray}}
\def\gapp{\, \raisebox{-.5ex}{$\stackrel{>}{\scriptstyle\sim}$}\, }
\def\lapp{\, \raisebox{-.5ex}{$\stackrel{<}{\scriptstyle\sim}$}\, }

\begin{document}

\title{HUNTING DOWN THE QUARK-GLUON PLASMA \\
       IN RELATIVISTIC HEAVY-ION COLLISIONS}

\author{ULRICH HEINZ$^{a,b}$}

\address{$^a$Theory Division, CERN, CH-1211 Geneva 23, Switzerland\\
E-mail: Ulrich.Heinz@cern.ch\\
$^b$Institut f\"ur Theoretische Physik, Universit\"at Regensburg,\\
D-93040 Regensburg, Germany} 

\maketitle\abstracts{The present status of the heavy-ion program to
  search for quark-gluon plasma is reviewed. The goal of this program
  is to recreate the Big Bang in the laboratory, by generating small
  chunks of exploding quark-gluon plasma (``The Little Bang''). I
  argue that the analogues of the three pillars of Big Bang Theory
  (Hubble flow, microwave background radiation, and primordial
  nucleosynthesis) have now been firmly established in heavy-ion 
  collisions at SPS energies: there is convincing evidence for strong
  radial flow, thermal hadron emission, and primordial hadrosynthesis
  from a color-deconfined initial stage. Direct observation of the 
  quark-gluon plasma phase via its electromagnetic radiation will be
  possible in planned collider experiments at higher energies. 
}

%%%%%%%%%%%%%%%%%%%%%%%%%%%%%%%%%%%%%%%%%%%%%%%%%%%%%%%%%%%%%%%%%%%%%%%
\section{Introduction: What Are We Looking For?}
\label{sec1}
%%%%%%%%%%%%%%%%%%%%%%%%%%%%%%%%%%%%%%%%%%%%%%%%%%%%%%%%%%%%%%%%%%%%%%%

The highest particle and energy densities existed for a fleeting
moment in the early history of our universe, shortly after the 
``Big Bang''. In heavy-ion collisions at ultrarelativistic energies
one hopes to be able to recreate such high-density matter and to study
its properties. At presently available beam energies the goal is to
pass the energy density threshold for color-deconfinement\,\cite{Karsch},
$\epsilon_{\rm cr} \lapp 1$ GeV/fm$^3$. If such energy densities can 
be reached and the energy sufficiently thermalized, strongly 
interacting matter will manifest itself as a {\bf quark-gluon plasma}: 
the hadronic constituents (quarks and gluons) become deconfined, and 
hadrons loose their identity.

%%%%%%%%%%%%%%%%%%%%%%%%%%%%%%%%%%%%%%%%%%%%%%%%%%%%%%%%%%%%%%%%%%%%%%%
\subsection{The Big Bang model as an example}
\label{sec1.1}
%%%%%%%%%%%%%%%%%%%%%%%%%%%%%%%%%%%%%%%%%%%%%%%%%%%%%%%%%%%%%%%%%%%%%%%

How do we know that high energy densities existed in the Early
Universe? How is the Big Bang Theory ``proved''? It is worthwhile to
contemplate this question for a minute. According to the textbooks the 
Big Bang Theory rests on three pillars: (1) The Hubble law relating 
galaxy distances to their recession velocities, which proves that our 
universe is expanding; (2) the cosmic microwave background (CMB) with
its uniform temperature of 2.7~K; and (3) the successful prediction of
the primordial abundances of small atomic nuclei during nucleosynthesis. 
The present {\bf expansion rate} of our universe reflects its 
{\bf initial conditions} and the {\bf equation of state} of the matter 
in the universe, integrated over its lifetime. 
{\em Primordial nucleosynthesis} happened when the universe had cooled 
down to about $T\simeq 100$ keV, about 3 minutes after the Big
Bang; this is the stage of {\bf chemical freeze-out} of our universe
at which its original chemical composition was fixed (later to be
modified inside stars by non-cosmological mechanisms). The 
{\em microwave background radiation} signals the recombination of
atoms and the depletion of free charges in the early universe,
rendering it transparent to photons. This {\bf thermal freeze-out} of
our universe happened much later, about 400000 years after the Big
Bang when it had cooled down to 
$T\simeq {1\over 4}\,{\rm eV} \simeq 3000$ K.

Before thermal freeze-out, the universe was in nearly perfect local 
and global {\em thermal} equilibrium (except for the neutrinos which 
froze out earlier and today have a temperature of 1.95~K). Before the 
onset of nucleosynthesis, it was also in a state of {\em chemical} 
equilibrium. Thermodynamic equilibrium being a state of maximum 
entropy or minimal information, all memory about the earlier stages 
of the universe got wiped out, leaving practically no 
directly observable traces. In particular, the color-confining phase
transition with its formation of hadrons from quarks and gluons at
about 10 $\mu$s after the Big Bang, which we now want to study in the
laboratory, is in cosmology safely hidden behind the curtain of
thermal CMB photons. No {\bf direct probes} from the early stages are
known which survived this tendency for re-equilibration; given the
roughly 18 orders of magnitude between the (slow) expansion rate of
the universe and the (fast) strong, weak and electromagnetic
interaction rates among the particles this is, of course, not
surprising. At the moment our best bet for direct signals from the
very early universe seems to be a detailed and quantitative analysis 
of the {\bf fluctuations} in the CMB spectrum which occur at the level 
of about 10$^{-5}$ of the average signal.

Still, with only a single observed event and little or no direct
experimental evidence about its very early stage to show, the Big Bang
Theory has become the almost unanimously accepted model for the
evolution of our cosmos. How can this be explained? The most important
reason is, I think, the fact that we can analyze our present state (the 
{\bf final state} of the ``Big Bang experiment'') by making exceedingly
accurate observations. This provides stringent constraints for the
extrapolation backwards in time. Furthermore, general relativity,
which has been tested elsewhere, provides a robust and highly reliable
theory for the dynamical evolution of the cosmos, given the equation
of state of the matter. And the latter again is known experimentally
and theoretically to be simple during most parts of the dynamical
evolution, with the exception of a few critical stages (typically near
phase transitions) which correspondingly draw most of the present
theoretical attention.   

%%%%%%%%%%%%%%%%%%%%%%%%%%%%%%%%%%%%%%%%%%%%%%%%%%%%%%%%%%%%%%%%%%%%%%%
\subsection{What is a quark-gluon plasma?}
\label{sec1.2}
%%%%%%%%%%%%%%%%%%%%%%%%%%%%%%%%%%%%%%%%%%%%%%%%%%%%%%%%%%%%%%%%%%%%%%%

The initiated reader will not have missed the strong parallelism
between the above discussion of the Big Bang and the field of
relativistic heavy-ion collisions. Very similar concepts (indicated 
above by emphasized key-words) play a crucial role on both contexts. 
Why, then, do we not have a generally accepted theory for the 
``Little Bang''? 

The main reason is the small geometric size of the object under
study. Contrary to the Early Universe, the region of high density is
limited to a region of at most a few 10 fm in diameter, surrounded by
vacuum. Once the energy in the reaction zone begins to thermalize, it
builds up pressure which leads to explosive expansion of the
fireball. The associated expansion time scale is no longer 18 orders
of magnitude, but only perhaps a factor 10 to 100 larger than that of
the microscopic equilibration processes. This introduces basic
uncertainties into dynamical evolution models which to eliminate
requires hard work and many detailed tests. It is fair to say that the
goal of a generally accepted ``standard model of nuclear fireball
dynamics'' has not yet been reached. 

Things would be much simpler if, as in cosmology, we were sure that
hydrodynamics can be used. In fact, the conditions for creating a
quark-gluon plasma (QGP) and for being able to describe its dynamics
hydrodynamically are very closely related. Both involve an assessment
of the degree of local equilibrium that can be reached. I would list 
the following criteria for being permitted to call a small, short-lived, 
rapidly expanding system of quarks and gluons a quark-gluon {\bf plasma}:

\begin{enumerate}

\item
Sufficient {\em energy density} for deconfinement, $\epsilon \gg
\epsilon_{\rm cr} \simeq 1$ GeV/fm$^3$.

\item   
Locally {\em thermal momentum distributions} with the same temperature
$T$ for all particle species, in a limited momentum range $0{<}p\lapp 6\,T$. 
To establish collective plasma properties and to validate a hydrodynamic 
approach, it is not necessary to fully equilibrate the tails of the 
momentum distributions. In fact, given the short lifetime and small 
size of the reaction zone, some fast particles produced by hard QCD 
processes will always escape, giving a power-law tail to the momentum 
spectrum. Note that {\em chemical equilibrium} is also not necessary: 
the relative abundances of gluons and light and heavy (anti)quarks need 
not be equilibrated for plasma properties and hydrodynamic behavior to 
manifest themselves. Chemical non-equilibrium requires, however, the
introduction of additional chemical potentials in the equation of
state.

\item
Sufficiently {\em large volume} and {\em strong rescattering}. This
and the following criterium are the most subtle ones: the relevant
volume is characterized by the mean free path which again reflects the 
scattering rate. To ensure thermalization by kinetic equilibration we
must require 
 \be
 \label{1}
   {V_{\rm hom}\over \lambda_{\rm mfp}^3} \gg 1 
 \ee
where in perturbation theory the mean free path is given\,\cite{HW96,Yaffe} 
in terms of the time-dependent temperature $T(\tau)\sim\epsilon(\tau)^{1/4}$ 
by\footnote{This implies a transport cross section $\sigma_{\rm trans} 
  \sim \al^2 \ln(1/\al)/T^2$ which grows as $T$ decreases.}
 \be
 \label{2}
   \lambda_{\rm mfp}(\tau) = \tau_{\rm scatt}(\tau)\, \langle v\rangle
   \sim {1\over {\#}\,T(\tau)\,\al^2(\tau)\ln(1/\al(\tau))}
 \ee
(with ${\#}$ indicating a number proportional to the number of degrees 
of freedom in the plasma), and $V_{\rm hom}(p)$ is the ``homogeneity
volume'' of particles with momentum $p$:\,\cite{WH99} 
 \be
 \label{3}
   V_{\rm hom}(m_\perp,y,\tau) = 2\pi\, R_\parallel(m_\perp,y,\tau)\,
   R_\perp^2(m_\perp,y,\tau).
 \ee
$R_\parallel,R_\perp$ are the longitudinal and transverse
``homogeneity lengths'' as measured by Bose-Einstein correlations.
For thermalized expanding fireballs they can be estimated by simple
analytical expressions\,\cite{WH99} which, assuming boost-invariant
longitudinal expansion,\cite{Bj} lead to
 \bea
 \label{4}
   V_{\rm hom}(m_\perp,y,\tau) &=& {V_{\rm geom}(\tau) \over
   \left( 1 + \eta_f^2(\tau) {m_\perp\over T}\right)
   \sqrt{1+ (\Delta\eta)^2(\tau) {m_\perp\over T}}}
 \nonumber\\
   &&\longrightarrow \ \propto V_{\rm geom}(\tau)  
   \left( {T(\tau)\over m_\perp}\right)^{3/2}\, .
 \eea
Here the limit is for large $m_\perp$, $\eta_f(\tau)$ is the average 
transverse flow rapidity of the fireball, $\Delta\eta(\tau)$ its
longitudinal extension in space-time rapidity, and $V_{\rm
  geom}(\tau)=\pi R^2(\tau)\cdot 2\tau \Delta\eta(\tau)$ is the
geometric fireball size (growing with time). The interesting
ratio (\ref{1}) then becomes
 \be
 \label{5}
   {V_{\rm hom}(m_\perp,\tau)\over \lambda_{\rm mfp}^3(\tau)} \approx
   V_{\rm geom}\, (\tau)T^3(\tau) \cdot {\#}^3
     \al^6(\tau)\ln^3(1/\al(\tau))
   \left( {T(\tau)\over m_\perp}\right)^{3/2}\, .
 \ee
Entropy conservation requires the first factor to be time-independent
whereas the second factor grows with time. One sees that the criterium 
(\ref{1}) is the more easy to fulfill the higher the initial
temperature ($T_0=T(\tau_0)$), the larger the number of massless degrees of
freedom in the plasma (${\#}$), the more strongly they interact
($\al$), and the smaller their transverse mass ($m_\perp$). Particles
with large $m_\perp$ will always be difficult to thermalize. 

\item
Sufficiently {\em long lifetime}:
 \be
 \label{6}
   \tau_{\rm scatt} \simeq {\lambda_{\rm mfp} \over c} \ll
   \theta^{-1} = {1\over \partial\cdot u} = \tau_{\rm exp} .
 \ee
The divergence of the flow velocity field $\theta(x) = \partial \cdot
u(x)$ gives the local expansion rate in the fireball; it is the analog
of the Hubble constant in cosmology. For $d$-dimensional scaling
expansion it is given by $\theta^{-1}=\tau_{\rm exp} = \tau/d$ whereas
the temperature drops like $T(\tau) = T_0 (\tau_0/\tau)^{d/3}$, with
$T_0\tau_0\approx 1$.\cite{HW96,Bj} Putting this together we find
 \be
 \label{7}
   {\tau_{\rm scatt}\over\tau_{\rm exp}}
   = {d\, (T_0\tau)^{d/3-1}
      \over {\#} \al^2(\tau)\ln(1/\al(\tau))}\, .
 \ee
For $d$=1 the expansion looses and the conditions for thermalization 
get better and better with time; the case $d$=3 is marginal\,\cite{HW96}, 
and only the increasing coupling strength $\al(\tau)$ at lower 
temperatures works in favor of increased thermalization.\cite{Wong} 
Of course, after hadronization the cross sections become 
$\tau$-independent while the particle density continues to decrease 
such that eventually thermalization stops and the particle momenta 
freeze out.

\end{enumerate}

\noindent
Clearly, the last two criteria require sufficiently strong coupling 
$\al$. Small values of $\al$ are certainly disadvantageous for 
thermalization in heavy-ion collisions; fortunately, in real life 
$\al$ seems to be sufficiently large (much to the dismay of the 
practitioners of perturbative QCD).

%%%%%%%%%%%%%%%%%%%%%%%%%%%%%%%%%%%%%%%%%%%%%%%%%%%%%%%%%%%%%%%%%%%%%%%
\section{Heavy-Ion Observables: Hard and Soft Probes}
\label{sec2}
%%%%%%%%%%%%%%%%%%%%%%%%%%%%%%%%%%%%%%%%%%%%%%%%%%%%%%%%%%%%%%%%%%%%%%%

The observables in relativistic heavy-ion collisions can be divided
into two classes: hard and soft probes. {\em Hard} probes are created
early in the collision and, due to the {\em finite size} of the reaction
fireball and their relatively small reinteraction cross section, they
decouple early. They include hard direct photons and lepton pairs, 
hadronic jets, and charmonia ($J/\psi,\psi',\dots$). {\em Soft} probes 
are the light quark flavors $u,d,s$ and the corresponding hadrons, as 
well as the soft electromagnetic radiation emitted by them. They are 
created throughout the collision history, and the strongly interacting 
ones decouple late, triggered by the {\em expansion} of the fireball rather 
than by its finite size. Soft probes include the hadron abundance ratios, 
their momentum spectra and their momentum correlations.

The soft probes contain equivalent information as the known signatures of
the cosmological Big Bang: Hadron abundancies give information about the
{\em chemical equilibration} time scales; here strangeness plays a crucial 
role since the time scale $\tau_{\bar s s}$ for creating strange quark 
pairs is roughly of the same order of magnitude as the fireball lifetime
such that moderate changes in $\tau_{\bar s s}$ due to new physics (e.g. 
color deconfinement and the onset of gluon fusion\,\cite{RM82}) can cause 
major effects. The hadronic momentum spectra and two-particle correlations
provide information on {\em thermalization and collective flow}. Since flow
is generated by pressure, the observed flow pattern in the final state
gives a time integral of the pressure history in the fireball (and thus 
its equation of state), with different types of flow (radial expansion 
in central collisions, directed and elliptic flow in semiperipheral ones)
receiving different weights from early and late dynamical stages.\cite{S98} 
The observed flow patterns are thus the ``Little Bang'' analogues of 
the Hubble expansion in cosmology.

The Little Bang can therefore be reconstructed from the soft probes 
in very much the same way as the cosmological Big Bang; hard probes,
which escape directly from the early stages of the Little Bang and for
which no cosmological analogue exists (due to the absence of spatial 
boundaries of the universe), can then be used to check the consistency
of the reconstruction. The presently available data allow for the
realization of the first part of this program; the second part will 
require the higher collision energies becoming available at RHIC and LHC 
in the next few years. 

%%%%%%%%%%%%%%%%%%%%%%%%%%%%%%%%%%%%%%%%%%%%%%%%%%%%%%%%%%%%%%%%%%%%%%%
\section{Hard Probes: The Present Status}
\label{sec3}
%%%%%%%%%%%%%%%%%%%%%%%%%%%%%%%%%%%%%%%%%%%%%%%%%%%%%%%%%%%%%%%%%%%%%%%

An extensive review of the present status of QGP signatures has just 
appeared.\cite{Bass} It contains a comprehensive list of original 
references to which I refer the reader for details. 

Direct photons of other than hadronic decay origin have been searched 
for by the WA80, WA98 and CERES/NA45 collaborations at the CERN SPS
in S+Au and Pb+Pb collisions. So far only an upper limit of about 5-7\% 
above hadronic decay background could be established in S+Au collisions; 
the analysis of Pb+Pb data is still in progress. This eliminates
hydrodynamic expansion models with equations of state which do not
include a phase transition to QGP or at least a strong softening of 
the equation of state by resonance and string excitations at high 
temperatures. A positive signal of thermal QGP radiation (whose rate 
grows with $T^4$) is, however, only expected at higher beam energies 
where larger initial temperatures can be reached. Whether it can be 
extracted from the hadronic background depends on the unknown transverse 
momentum distribution of the latter\,\cite{Vesa} (which is affected 
by collective expansion).

An excess of dimuons in the mass region below the $J/\psi$ has been 
reported, in one form or another, by the HELIOS3, NA38 and NA50 
collaborations in S- and Pb-induced heavy-ion collisions. This excess 
appears to have a non-linear dependence on the charged hadron 
multiplicity (collision centrality). Its origin is presently
unclear.\cite{Scomp}   

The CERES/NA45 collaboration has seen in S+Au and Pb+Au collisions
a strong excess of $e^+e^-$ pairs with masses 200 MeV$<m_{ee}<m_\rho$, 
with a non-linear multiplicity dependence and concentrated at low 
transverse momentum.\cite{CERES} It can probably be explained by 
%
%%%%%%%%%%%%%%%%%%%%%%%%%% Fig. 1 %%%%%%%%%%%%%%%%%%%%%%%%%%%%%%%%%%%%%%%
\vspace*{-1truecm}
\begin{figure}[ht]
\centerline{\epsfxsize=11cm\epsffile{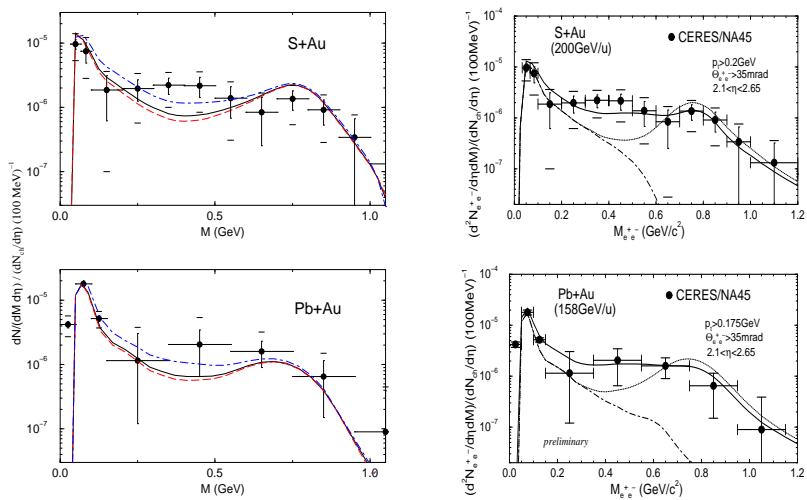}}
\vspace*{-1.2truecm}
\caption{CERES dilepton mass spectra\,\protect\cite{CERES} and theoretical
  simulations with an expanding fireball model, using different 
  theoretical approaches\,\protect\cite{Wambach} to the in-medium dilepton 
  rates.
\label{F1}}
\end{figure}
%\vspace*{-0.2cm}
%%%%%%%%%%%%%%%%%%%%%%%%%%%%%%%%%%%%%%%%%%%%%%%%%%%%%%%%%%%%%%%%%%%%%%%%%
%
hadronic mechanisms, but requires a very dense and hot hadronic medium 
in which pions and $\rho$-mesons violently rescatter, leading to a 
broadening of the vector meson spectral densities\,\cite{Wambach} 
(Fig.~\ref{F1}).

Jet production and jet quenching by the dense fireball medium\,\cite{Wang}
cannot be studied at present SPS energies due to insufficient rates, but 
will become accessible at RHIC.

\begin{figure}[ht]
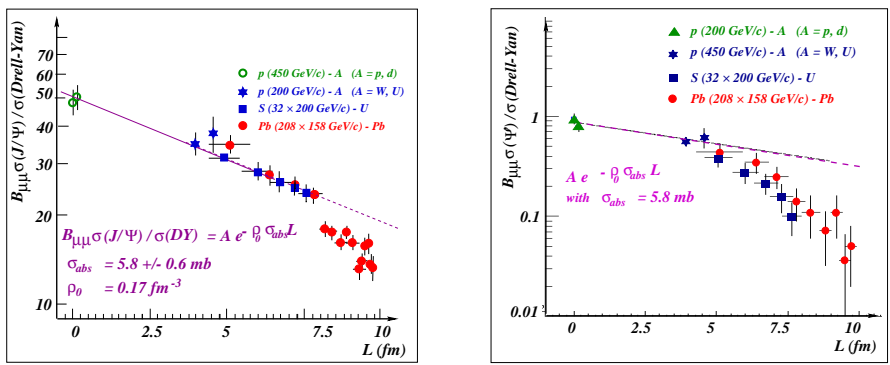

\begin{center}
   \begin{minipage}[t]{5.5truecm}
         \epsfxsize 5.5truecm \epsfbox{Fig2a.eps}
         \hfill
   \end{minipage}
   \hfill  
   \begin{minipage}[t]{5.5truecm}
         \epsfxsize 5.30truecm \epsfbox{Fig2b.eps}
         \hfill
   \end{minipage}
\end{center}
\vspace*{-0.5cm}
\caption{Preliminary data on $(J/\psi)/DY$ and $\psi'/DY$ from the 
  NA38/NA50 collaboration\,\protect\cite{NA50}, plotted as a function 
  of the Glauber model nuclear thickness parameter $L$ (see text). 
\label{F2}}
\end{figure}

Much recent attention focussed on the discovery by NA38/NA50 of 
``ano\-malous'' suppression of $J/\psi$ and $\psi'$ vector mesons in 
central nucleus-nucleus collisions. Different collision systems 
can be compared by introducing a theo\-re\-ti\-cal auxiliary variable, 
the average path length $L$ which the ${\bar c c}$-pair
must traverse before escaping the reaction zone.\cite{NA50} ``Normal''
suppression is defined by an exponential attenuation 
$B\sigma_\psi/\sigma_{DY} \propto \exp(-\rho\sigma_{\rm abs}L)$ 
(straight line in Fig.~\ref{F2}) with the normal nuclear density $\rho$ 
and a fitted $\psi N$ absorption cross section $\sigma_{\rm abs}$.
A deviation from this behaviour is first observed for the $\psi'$ near
$L=5$ fm, then for the $J/\psi$ near $L=7.5$ fm. Since about 32-40\% 
(5-8\%) of the observed $J/\psi$ stem from radiative $\chi_c$ ($\psi'$) 
decays, it is not unlikely that the drop near $L=7.5$ fm indicates 
anomalous $\chi_c$ suppression, and that anomalous suppression of the 
$J/\psi$ itself requires even larger values of $L$. It is interesting 
that this supression pattern follows the binding energies of the 
corresponding charmonium states: suppression of the more strongly bound 
states requires a higher density and/or lifetime of the fireball, here 
parametrized (perhaps not very fortunately) via $L$. 
%
%%%%%%%%%%%%%%%%%%%%%%%%%% Fig. 2 %%%%%%%%%%%%%%%%%%%%%%%%%%%%%%%%%%%%%%%
\vspace*{-0.3cm}
\vspace*{-0.2cm}
%%%%%%%%%%%%%%%%%%%%%%%%%%%%%%%%%%%%%%%%%%%%%%%%%%%%%%%%%%%%%%%%%%%%%%%%
%
At present the suppression mechanism is not fully understood 
theoretically,\cite{Kharzeev} but one condition appears to be 
unavoidable: strong rescattering of the $\bar c c$-pair in a 
{\em very dense environment}, probably of partonic origin. A 
consistent quantum mechanical and dynamical theory of this 
phenomenon is urgently needed.

%%%%%%%%%%%%%%%%%%%%%%%%%%%%%%%%%%%%%%%%%%%%%%%%%%%%%%%%%%%%%%%%%%%%%%%
\vspace*{-0.2truecm}
\section{Soft Probes: Hadronization, Thermalization and Flow}
\label{sec4}
%%%%%%%%%%%%%%%%%%%%%%%%%%%%%%%%%%%%%%%%%%%%%%%%%%%%%%%%%%%%%%%%%%%%%%%

Soft hadrons, which constitute the bulk of the produced particles,
provide a much richer body of experimental information which by now
has been analyzed in considerable quantitative detail. As I will show,
they give convincing evidence that we have seen ``the Little Bang''.

%%%%%%%%%%%%%%%%%%%%%%%%%%%%%%%%%%%%%%%%%%%%%%%%%%%%%%%%%%%%%%%%%%%%%%%
\subsection{Primordial hadrosynthesis}
\label{sec4.1}
%%%%%%%%%%%%%%%%%%%%%%%%%%%%%%%%%%%%%%%%%%%%%%%%%%%%%%%%%%%%%%%%%%%%%%%

\begin{figure}[ht]
\centerline{\epsfxsize=8cm\epsffile{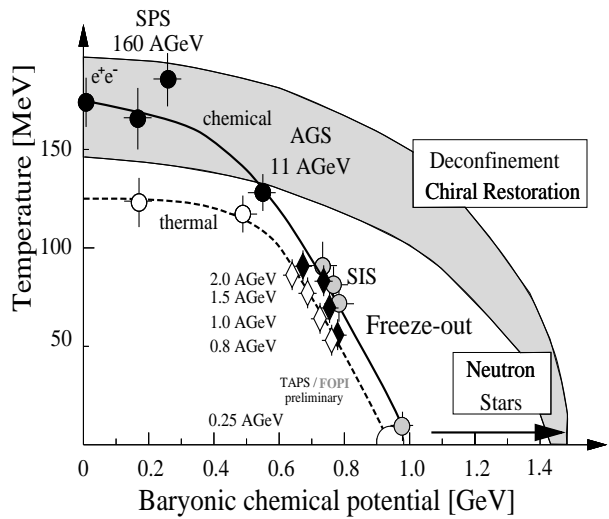}}
\caption{Compilation of freeze-out points in $e^+e^-$ collisions and
in heavy-ion collisions from SIS to SPS energies. Filled symbols: 
chemical freeze-out points from hadron abundances. Open symbols: 
thermal freeze-out points from momentum spectra and two-particle 
correlations. The shaded region indicates the parameter range of 
the expected transition to a QGP.
\label{F3}}
\end{figure}

It has been known for many years that hadron production in high energy 
physics exhibists striking statistical features. Recently several
rather detailed analyses were performed of the relative yields of 
different hadronic species produced in $e^+e^-$, $pp$, $p\bar p$ 
and $AA$ collisions, with the question in mind to what extent the 
hadronic final state reflects a state of {\em chemical 
equilibrium}.\cite{SQM98} The result is shown in Fig.~\ref{F3}.
%
%%%%%%%%%%%%%%%%%%%%%%%%%% Fig. 3 %%%%%%%%%%%%%%%%%%%%%%%%%%%%%%%%%%%%%%%%
\vspace*{-0.3cm}
%%%%%%%%%%%%%%%%%%%%%%%%%%%%%%%%%%%%%%%%%%%%%%%%%%%%%%%%%%%%%%%%%%%%%%%%%%
%

The single $e^+e^-$ point represents a large number of $e^+e^-$, 
$pp$ and $\bar p p$ collision systems at different center of mass 
energies 20 GeV\,$\leq \sqrt{s}\leq$\,900 GeV which were all 
found\,\cite{Becattini} to reflect chemical equilibrium hadron 
abundancies at a universal hadronization temperature 
$T_{\rm had}\approx 175$ MeV. The only clear deviation from 
chemical equilibrium in these systems is an undersaturation of 
overall strangeness, reflected in the ratio of produced strange 
to non-strange quark pairs $\lambda_s = 2\langle \bar s s\rangle/
(\langle\bar u u \rangle + \langle\bar d d\rangle)|_{\rm produced} 
\approx 0.2{-}0.25$, again almost independent of 
$\sqrt{s}$.\cite{Becattini,BGS98}. But even if the total number of 
$\bar s s$ valence quark pairs is below the chemical equilibrium
value, the available $s$ and $\bar s$ quarks are distributed among 
the various strange hadron species according to the law of maximum 
entropy. 

Given the small size of these collision systems it is impossible 
to imagine that this can be the result of equilibration by hadronic 
rescattering. It must reflect {\em pre-established} statistical 
equilibrium, i.e. the statistical filling of the available hadronic 
phase-space according to the law of Maximum Entropy at the point of 
hadron formation. The temperature $T_{\rm had}$ should be understood 
as a Lagrange multiplier for the energy density at which this happens;
its universality and numerical value tell us that hadronization 
always happens at the same critical energy density $\epsilon_{\rm cr}
\approx 1$ GeV/fm$^3$. At higher collision energies hadrons are not 
produced at higher temperatures, but over a {\em larger volume} at the 
{\em same energy density}.\cite{Becattini}. After formation
the hadrons do {\em not} rescatter, and the observed hadron abundances
thus reflect the {\em primordial} values established at the point of 
hadronization.

It is interesting that in heavy-ion collisions at the SPS (S+S, S+Ag, 
Pb+Pb) one finds again chemical equilibrium hadron abundances at the same 
chemical freeze-out temperature of about 180 MeV.\cite{BGS98} This suggests
hadron formation by the same statistical phase-space filling mechanism
%
%%%%%%%%%%%%%%%%%%%%%%%%%% Fig. 4 %%%%%%%%%%%%%%%%%%%%%%%%%%%%%%%%%%%%%%%%
\vspace*{-1.2truecm}
\begin{figure}[ht]
\centerline{\epsfxsize=6cm\epsffile{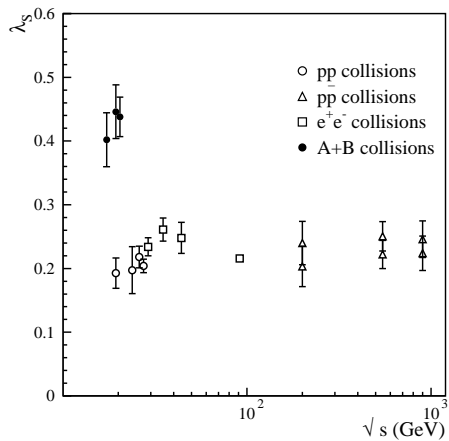}}
\vspace*{-1.6truecm}
\caption{The strangeness suppression factor $\lambda_s$ as a function of
$\protect\sqrt{s}$.\protect\cite{BGS98}. The two points each for $p\bar p$
collisions reflect the inclusion (exclusion) of the initial valence quarks.
\label{F4}}
\end{figure}
\vspace*{-0.2truecm}
%%%%%%%%%%%%%%%%%%%%%%%%%%%%%%%%%%%%%%%%%%%%%%%%%%%%%%%%%%%%%%%%%%%%%%%%%%
%
as in $e^+e^-$ and $pp$ collisions (reflecting the Maximum Entropy 
Principle), followed by immediate freeze-out of the hadron yields. Since
changes in the relative abundances require inelastic collisions whose
cross sections are mostly quite small, this is not unexpected, especially
if the fireball features collective expansion (which is known to foster 
freeze-out) already at hadronization. 

What is different from elementary collisions is that in heavy-ion
collisions a considerably larger global strangeness fraction
$\lambda_s\approx 0.4-0.45$ is measured (Fig.~\ref{F4}).\cite{BGS98} 
Following the above argument and realizing that hadronic 
{\em strangeness-creating} cross sections are particularly 
small,\cite{KMR86} this can only be due to processes {\em before} 
hadronization. Indeed, all known microscopic kinetic models based
entirely on $pp$ input and hadronic reinteraction dynamics
fail to reproduce this {\em strangeness enhancement}.\cite{Odyn}  
Heavy-ion reactions thus generate a {\em prehadronic state} with 
different dynamics (presumably a longer lifetime) than in $e^+e^-$ 
and $pp$ collisions. The factor 2 increase in the strangeness 
fraction $\lambda_s$ indicates a short strangeness saturation 
time scale $\tau_{s\bar s}$ in this state, as predicted for a 
QGP.\cite{RM82} In fact, Sollfrank\,\cite{SBRS98} has 
argued that the observed strangeness enhancement in $A+A$ collisions 
at the SPS is consistent with the hadronization of a 
{\em fully equilibrated} QGP at the hadronization temperature
$T_{\rm had}$ if the hadronization process itself leaves the 
strangeness/entropy ratio unchanged.

%
%%%%%%%%%%%%%%%%%%%%%%%%%%% Fig. 5 %%%%%%%%%%%%%%%%%%%%%%%%%%%%%%%%%%%%%%%%%%
\vspace*{-0.5cm}
\begin{figure}[ht]
\centerline{\epsfxsize=10cm\epsffile{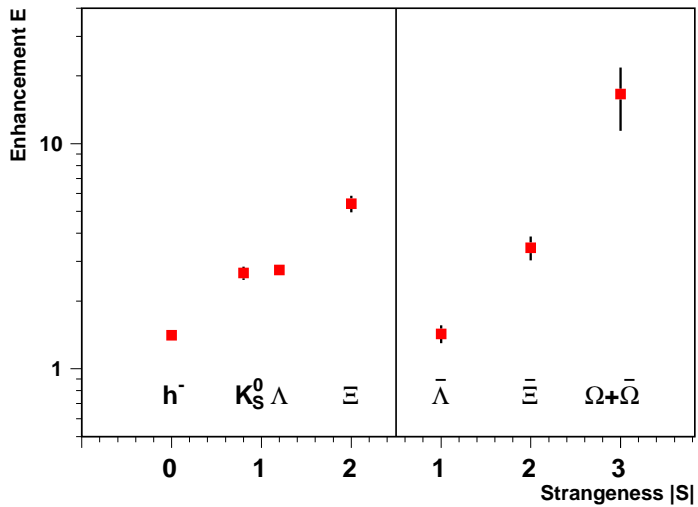}}
\vspace*{-0.3cm}
\caption{Enhancement factor for the mid-rapidity yields per participating 
nucleon in Pb+Pb relative to p+Pb collisions for various strange and 
non-strange hadron species.\protect\cite{WA97}
\label{F5}}
\end{figure}
%%%%%%%%%%%%%%%%%%%%%%%%%%%%%%%%%%%%%%%%%%%%%%%%%%%%%%%%%%%%%%%%%%%%%%%%%%
%
The global strangeness enhancement reflected in $\lambda_s$ occurs 
already in S+S collisions and remains roughly unchanged in Pb+Pb 
collisions.\cite{BGS98} A recent comparison of p+Pb and Pb+Pb collisions 
at the SPS\,\cite{WA97} (see Fig.~\ref{F5}) further shows that
while the bulk of the strangeness enhancement is carried by the kaons 
and hyperons ($\Lambda$, $\Sigma$), which are enhanced by about a factor 
3 near midrapidity, the enhancement is much stronger for the doubly and 
triply strange baryons $\Xi$ and $\Omega$ and their antiparticles, 
with an enhancement factor of about 17 (!) for $\Omega+\bar\Omega$ at 
midrapidity. Such a rise of the enhancement with the strangeness content 
of the hadrons is natural in statistical hadronization mo\-dels, but 
contradicts expectations based on the respective production thresholds 
in hadronic (re)interactions. WA97 also found\,\cite{WA97} that all of 
the enhancement factors in Pb+Pb are independent of the collision 
centrality, i.e. of the size of the midrapidity source, from about 100 
to 400 participating nucleons, and that similar enhancement patterns 
are seen in S+S collisions.\cite{WA97} Thus, whatever causes the 
enhancement in central Pb+Pb collisions exists already in S+S collisions! 

The existence of a prehadronic stage without color confinement, both in 
S+S and Pb+Pb collisions at the SPS, is supported by a recent argument 
by Bialas\,\cite{Bialas} which generalizes similar ideas by 
Rafelski\,\cite{R91} by removing the assumption of thermal equilibrium.
Bialas points out that by considering baryon/antibaryon production 
ratios the unknown baryon wave function drops out, and one can very 
easily test whether or not the baryons were formed by statistical 
hadronization (coalescence) of {\em uncorrelated} quarks. He finds
that the data from S+S and Pb+Pb, but not those from p+Pb follow
the corresponding simple ``quark counting rules''; the conspicuous 
absence of correlations among the quarks is interpreted in terms of 
a color-deconfined initial state in the nuclear collisions.\cite{Bialas}

While at the SPS chemical freeze-out appears to happen immediately at
hadronization (such that the measured hadron yields reflect the abundances
from the {\em primordial hadrosynthesis}), Fig.~\ref{F3} suggests that 
at the AGS and SIS chemical freeze-out occurs at lower temperatures, after 
some further ``chemical cooking''. This is probably due to longer 
lifetimes and slower expansion of the reaction zone at lower beam 
energies.

%%%%%%%%%%%%%%%%%%%%%%%%%%%%%%%%%%%%%%%%%%%%%%%%%%%%%%%%%%%%%%%%%%%%%%%
\subsection{Thermal hadron radiation and radial flow}
\label{sec4.2}
%%%%%%%%%%%%%%%%%%%%%%%%%%%%%%%%%%%%%%%%%%%%%%%%%%%%%%%%%%%%%%%%%%%%%%%

I now discuss the analogues of Hubble expansion and the cosmic
microwave background. In heavy-ion collisions the latter is not
generated by {\em photons} (because these decouple immediately after their
production, i.e. throughout the time evolution of the fireball), but
by {\em hadrons}. Due to their strong interactions, the fireball is
{\em opaque} to quarks and hadrons for most of its dynamical history,
and only at the end, after expansion has diluted the matter sufficiently,
the hadrons decouple. As in cosmology, where, after 400000
years of complete opaqueness, the universe became transparent to
photons quite suddenly after electron-ion recombination, this kinetic
decoupling process appears to happen in bulk and rather suddenly also
in heavy-ion collisions: two-particle Bose-Einstein correlation
measurements, which give access to the space-time structure of the
fireball at the point of freeze-out,\cite{WH99} indicate for Pb+Pb
collisions at the SPS an emission duration for pions (the most
abundantly produced particles) of not more than 2-3 fm/$c$, after a
total expansion time of at least 9-10 fm/$c$.\cite{NA49HBT,HJ99}
Consistently with that, most of the pions are emitted from the bulk of
the source and not only from a thin surface layer.\cite{TH98,T99} It
is also known\,\cite{SH94} that in heavy-ion collisions kinetic 
freeze-out of hadrons is triggered by expansion, not by the finite 
size of the source, again as in the early universe. 

What is, then, the temperature of this thermal hadron radiation? As
for the cosmic microwave background, it is determined from the 
energy spectrum of the radiated particles and, as in the consmological
context, this energy spectrum is affected by (Hubble) expansion.
Moreover, as in cosmology, the kinetic decoupling temperature reflected
in the energy or momentum spectra differs from the chemical freeze-out 
temperature reflected in the particle abundances; it is considerably
lower. 

Expansion flow affects the observed momentum spectra in two distinct
ways: since the source flows towards the detector, the single-particle
distributions are flattened by a blueshift effect,\cite{LHS90} and as
different parts of the expanding source recede from each other, the
homogeneity regions (Hanbury Brown/Twiss (HBT) size parameters)
measured by Bose-Einstein interfero\-metry (see Sect.~\ref{sec1.2}) are
reduced by velocity gradients in the source.\cite{WH99,HJ99} 

For the blueshift effect on single-particle transverse mass spectra
one must distinguish two domains: In the region of relativistic
momenta, $p_\perp{\gg}m_0$, the inverse slope $T_{\rm app}$
of all particle species is the same and given by the blueshift formula 
$T_{\rm app}{=}T_{\rm f} \sqrt{{1{+}\langle v_\perp\rangle \over 1{-}\langle
  v_\perp\rangle}}$. This formula does not allow to disentangle the
average radial flow velocity $\langle v_\perp\rangle$ and freeze-out
temperature $T_{\rm f}$. In the non-relativistic domain $p_\perp{\ll}m_0$
the inverse slope is given approximately by $T_{\rm app}{=}T_{\rm f}{+}m_0
\langle v_\perp^2 \rangle$, and the rest mass dependence of the
``apparent temperature'' (inverse slope) allows to determine $T_{\rm f}$ 
and $\langle v_\perp^2 \rangle$ separately.\cite{LHS90,NuXu} (In $pp$ 
collisions no $m_0$-de\-pen\-dence of $T_{\rm app}$ is seen.\cite{NuXu}) 

\begin{figure}[ht]
\begin{center}
   \begin{minipage}[t]{5.7truecm}
         \epsfxsize 5.7truecm \epsfbox{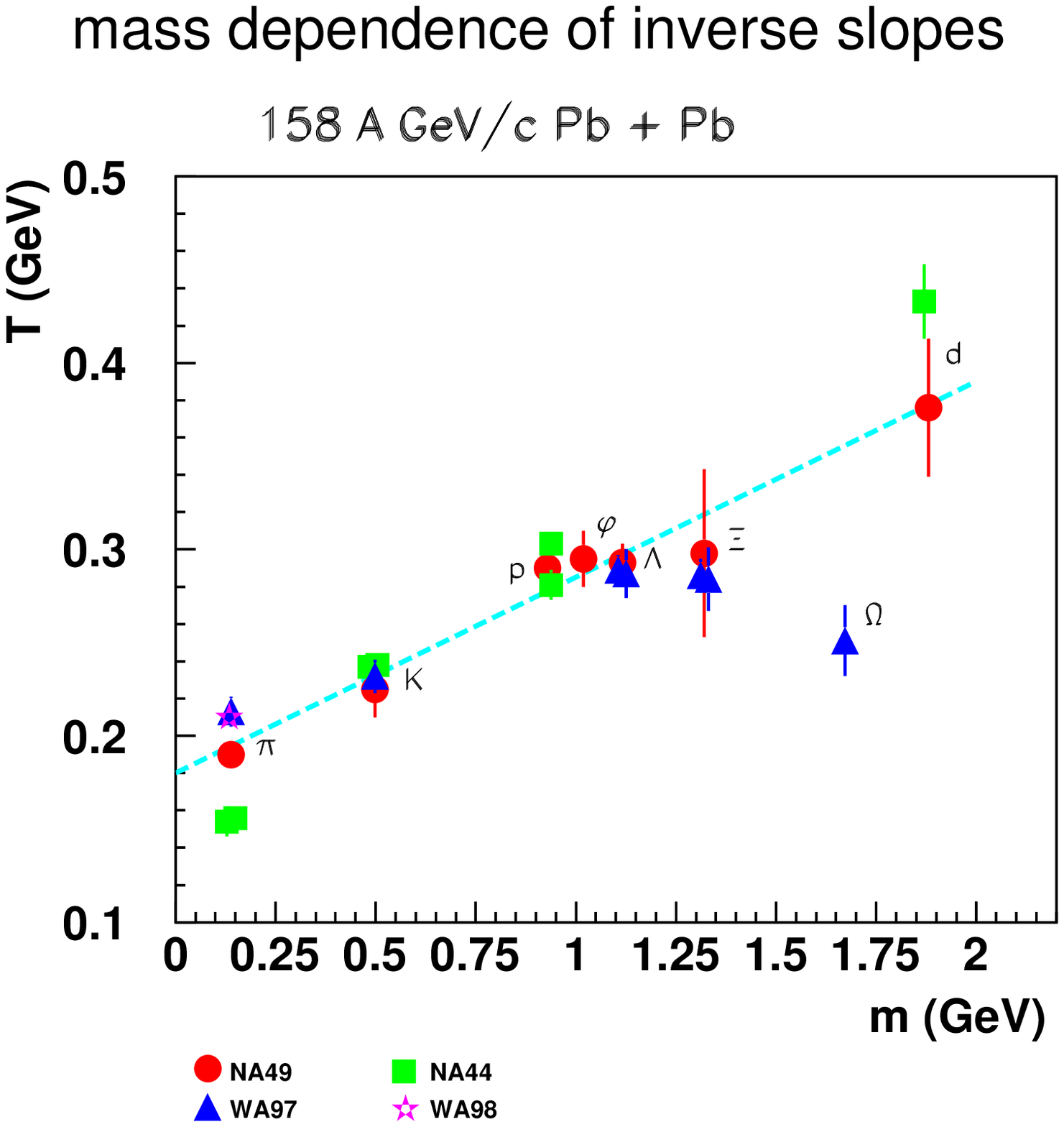}
         \hfill
   \end{minipage}
   \hfill  
   \begin{minipage}[b]{5.4truecm}
         \epsfxsize 5.4truecm \epsfbox{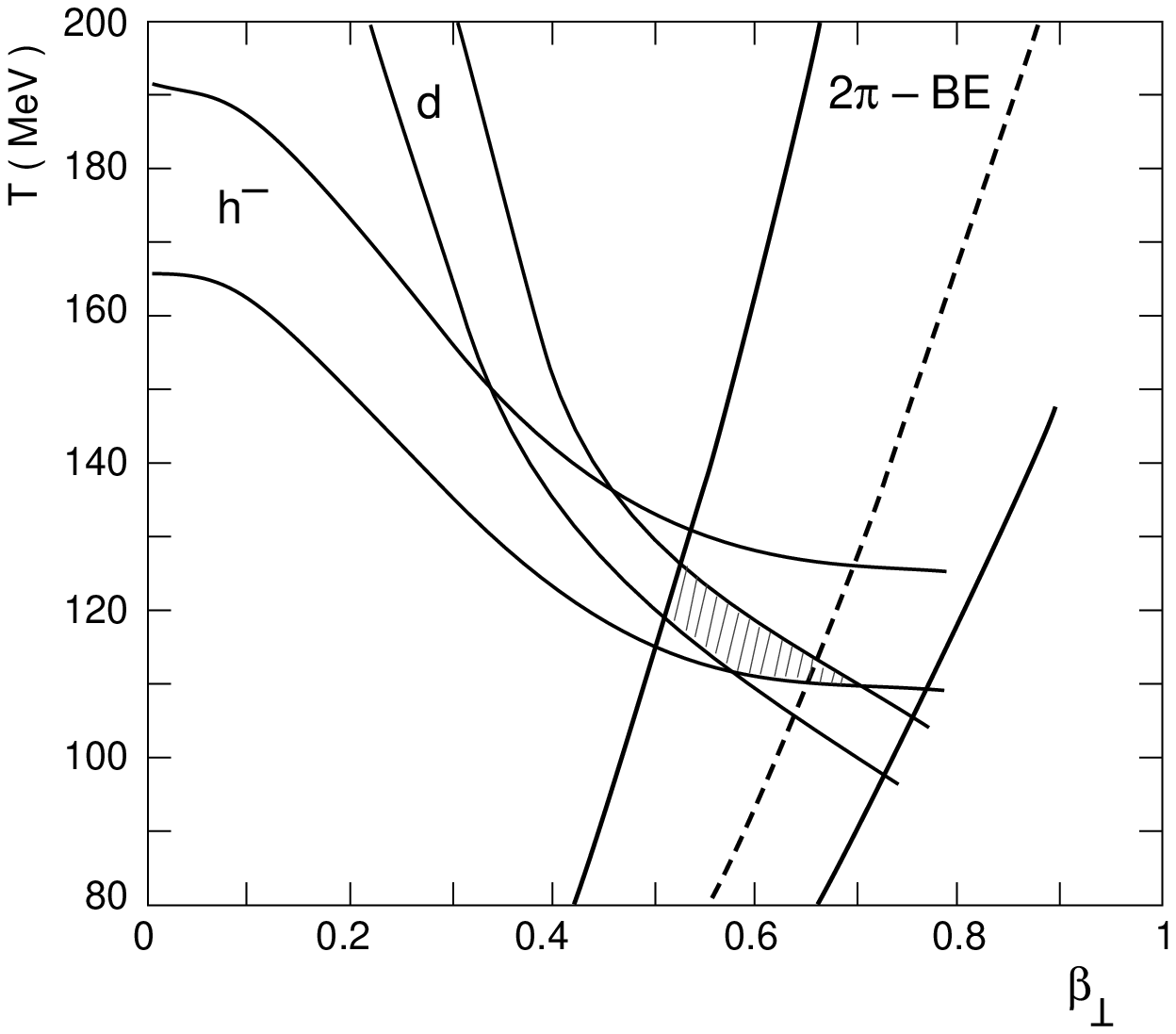}
         \hfill
   \end{minipage}
\end{center}
\vspace*{-0.5cm}
\caption{Left: Inverse transverse slope parameter ``$T$'' for
  different hadron species as a function of their rest masses,
  measured for Pb+Pb collisions at the SPS by various 
  experiments.\protect\cite{INPC98}
  Right: Constraints on possible combinations of kinetic freeze-out
  temperature $T$ and collective radial flow velocity $\beta_\perp$ 
  from the transverse momentum spectra of negative hadrons ($h^-$) and
  deuterons ($d$) and from the transverse momentum dependence of the
  transverse HBT radius parameter $R_\perp(m_\perp)$
  ($2\pi$-BE).\protect\cite{NA49HBT} 
\label{F6}}
\end{figure}

The left diagram in Fig.~\ref{F6} shows a compilation of measured
slope parameters $T_{\rm app}$ for a variety of hadron species in 
Pb+Pb collisions at the SPS. 
While a rise with the rest mass $m_0$ is clearly seen, providing
strong evidence for radial flow, some of the detailed features lead to 
ambiguities in the separation of temperature and flow. The pion slopes
are very sensitive to the $p_\perp$-region in which the fits are
performed, due to strong resonance decay contributions at low
$p_\perp$. Also, pions are always relativistic such that the formula
$T_{\rm app}{=}T_{\rm f}{+}m_0 \langle v_\perp^2 \rangle$ cannot be
used for them,\cite{Nix} and so they don't fit very well into the
systematics of that Figure. Finally, the $\Omega$ baryons (and perhaps
also the $\Xi$'s) exhibit steeper slopes than expected from this
formula. It was argued\,\cite{HSX98} that this reflects their earlier
kinetic freeze-out due to an absence of strong scattering resonances
with the dominating pions; these are essential for the kinetic
re-equilibration of the other hadron species.

A less ambiguous separation of temperature and flow is possible by
combining single-particle $m_\perp$-slopes with the pair-momentum
dependence of the transverse HBT radius extracted from 2-pion
correlations (right diagram in Fig.~\ref{F6}). 
%
%%%%%%%%%%%%%%%%%%%%%%%%%% Fig. 6 %%%%%%%%%%%%%%%%%%%%%%%%%%%%%%%%%%%%%%%
\vspace*{-0.3cm}
\vspace*{-0.1truecm}
%%%%%%%%%%%%%%%%%%%%%%%%%%%%%%%%%%%%%%%%%%%%%%%%%%%%%%%%%%%%%%%%%%%%%%%%
%
The latter is approximately given by\,\cite{WH99} 
$R_\perp(m_\perp)\approx R \left/
  \sqrt{1+\xi\langle v_\perp ^2\rangle {m_\perp\over T_{\rm f}}}\right.$
(see Eq.~(\ref{4})) where $m_\perp$ is the average transverse mass  
of the pions in the pair, $T_{\rm f}$ is the kinetic freeze-out
temperature, and $\xi$ is a number of order 1 which depends on the
detailed transverse velocity and density profiles. Since this formula
and the one for the inverse single-particle slopes provide orthogonal
correlations between $T_{\rm f}$ and $\langle v_\perp \rangle$ (see
Fig.~\ref{F6}), their combination leads to a rather accurate
separation of both. After proper translation of the symbols one
finds\,\cite{T99} for the average radial flow velocity $\langle
v_\perp \rangle\approx 0.5\,c$ and for the temperature of the thermal
hadron radiation $T_{\rm f}\approx 100$ MeV. (The somewhat lower
$T_{\rm f}$-value than in Fig.~\ref{F6} results from a more accurate
fit to the single-particle $h^-$ spectrum.\cite{T99}) 

Similar analyses were performed for smaller collision systems and
at lower beam energies; the results are shown by the open symbols in
Fig.~\ref{F3}. As in the Early Universe freeze-out of the momentum 
spectra is seen to happen later than the decoupling of particle 
abundances, at significantly lower temperatures. The ability of the 
system to remain in a state of approximately local thermal equilibrium 
while building up collective expansion flow and cooling was recently 
demonstrated in microscopic simulations with the URQMD model.\cite{Bravina}

%%%%%%%%%%%%%%%%%%%%%%%%%%%%%%%%%%%%%%%%%%%%%%%%%%%%%%%%%%%%%%%%%%%%%%%
\subsection{Initial energy density}
\label{sec4.3}
%%%%%%%%%%%%%%%%%%%%%%%%%%%%%%%%%%%%%%%%%%%%%%%%%%%%%%%%%%%%%%%%%%%%%%%

Two-particle Bose-Einstein correlation measurements give access to
both the geometry and collective dynamics of the fireball at the point
of kinetic freeze-out.\cite{WH99,HJ99} One not only finds signs of
strong radial flow of about 0.5\,$c$, as just discussed, but also
evidence for strong transverse {\em growth} of the fireball between
impact and freeze-out, by more than a factor 2.\cite{H97} The two
observations are dynamically consistent, given the lower limit on the
total expansion time $\tau_f\gapp 8$ fm/$c$ which can be
obtained\,\cite{HJ99,T99} from the longitudinal HBT radius
$R_\parallel$. Knowing the freeze-out temperature $T_{\rm f}$ and collective 
flow velocity, the thermal and collective flow energy density of the
system at freeze-out can be calculated. From the measured transverse
growth factor and the known longitudinal expansion pattern (all
extracted from single-particle spectra and HBT  measurements) one can
estimate the total geometric expansion factor of the fireball between
the onset of transverse expansion and freeze-out.\cite{H97}
Energy conservation then gives an estimate of the energy density at 
the beginning of transverse expansion, by multiplying the freeze-out
energy density with the volume expansion factor. 

This estimate has the advantage over the one using Bjorken's 
formula\,\cite{Bj} that both factors are determined more or less 
directly from (single-particle and correlkation) measurements, thus 
avoiding uncontrolled model assumptions (like the identification of 
momentum-space with coordinate-space rapidity densities) and free 
parameters (as the equilibration time $\tau_0$ in Bjorken's formula).
For Pb+Pb collisions at the SPS one finds\,\cite{H97} in this way 
initial energy densities of 2.5 -- 4 GeV/fm$^3$. This is comfortably 
above the critical energy density for deconfinement, consistent with 
the other arguments for an early pre-hadronic stage given above. Also, 
this energy density must have been at least partially thermalized 
because pressure (a consequence of thermalization) is necessary to 
drive the transverse expansion.   

%%%%%%%%%%%%%%%%%%%%%%%%%%%%%%%%%%%%%%%%%%%%%%%%%%%%%%%%%%%%%%%%%%%%%%%
\subsection{Elliptic flow: evidence for ``early pressure''}
\label{sec4.4}
%%%%%%%%%%%%%%%%%%%%%%%%%%%%%%%%%%%%%%%%%%%%%%%%%%%%%%%%%%%%%%%%%%%%%%%

In addition to radial transverse flow, which is typical for central
collisions, two other types of collective flow occur in collisions
with finite impact para\-me\-ter: in-plane directed flow (``bounce-off'') 
and elliptic flow.\cite{SG86,O93}. While the first of these is
concentrated at forward and backward rapidities, elliptic flow is
strongest at midrapidity. At very high energies the directed flow
becomes weak,\cite{Rischke} and only elliptic flow survives. Both
types of directed flow can be identified by a harmonic analysis of the
azimuthal dependence of the single-particle distributions around the
beam axis, with in-plane (elliptic) directed flow given by the first
(second) harmonic coefficient.\cite{O93}

Elliptic flow arises from the initial elliptic spatial deformation 
of the overlap region of the two colliding nuclei in the transverse
plane. If rescattering between secondaries created in this  
region builds up pressure sufficiently quickly, the resulting elliptic
anisotropy of the pressure gradients causes an elliptic deformation in 
the developing flow pattern: along the shorter dimension the pressure
gradient is larger and the flow develops faster. This quickly reduces
the geometric deformation, i.e. the developing elliptic flow
eliminates its own cause. This is the main reason why elliptic flow is 
particularly sensitive to the pressure and equation of state in the
{\em early} stages of the collision whereas radial flow receives strong
contributions also from the late stages.\cite{S98} 

%
%%%%%%%%%%%%%%%%%%%%%%%%%%% Fig. 7 %%%%%%%%%%%%%%%%%%%%%%%%%%%%%%%%%%%%%%%
\vspace*{-0.45truecm}
\begin{figure}[ht]
\centerline{\epsfxsize=12cm\epsffile{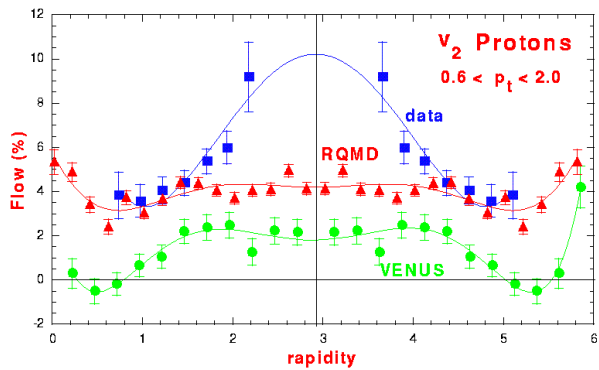}}
\vspace*{-2.5truecm}
\caption{Second harmonic coefficient $v_2$ of the azimuthal distribution 
  (elliptic flow) of protons in Pb+Pb collisions at the SPS, as a 
  function of rapidity. NA49 data\,\protect\cite{NA49flow} are shown
  together with RQMD and VENUS simulations.\protect\cite{Posk} 
\label{F7}}
\end{figure}
\vspace*{-0.1truecm}
%%%%%%%%%%%%%%%%%%%%%%%%%%%%%%%%%%%%%%%%%%%%%%%%%%%%%%%%%%%%%%%%%%%%%%%%%%
%
Elliptic flow was measured in Pb+Pb collisions at the SPS by the NA49
collaboration.\cite{NA49flow} Figure~\ref{F7} shows the data for
protons (due to their larger mass they show a stronger signal than
pions\,\cite{NA49flow}), together with simulations using RQMD and
VENUS event generators. Clearly the latter underpredict the effect
significantly. This failure is due to the particular way these codes
parametrize particle production: colliding nucleons create strings
which, after a formation time of about 1 fm/$c$, decay directly into
hadrons. During the initial stage of highest energy density the energy
is thus stored in non-interacting strings which do not contribute to
the pressure. In other words, their equation of state is
``ultra-soft''.\cite{S98} 

Sorge has recently shown\,\cite{S98} that a modification of the RQMD
code which simulates a harder QGP equation of state during the early
stage is able to generate the elliptic flow signal. Moreover, he
found that the {\em entire} effect is created during the high density
stage with $\epsilon > \epsilon_{\rm cr}$; after hadronization the
elliptic flow quickly saturates.\cite{S98} The elliptic flow 
data\,\cite{NA49flow} can thus be viewed as a rather direct glimpse of
the quark-gluon plasma.

%%%%%%%%%%%%%%%%%%%%%%%%%%%%%%%%%%%%%%%%%%%%%%%%%%%%%%%%%%%%%%%%%%%%%%%
\subsection{Absence of non-statistical event-by-event fluctuations}
\label{sec4.5}
%%%%%%%%%%%%%%%%%%%%%%%%%%%%%%%%%%%%%%%%%%%%%%%%%%%%%%%%%%%%%%%%%%%%%%%

In the introduction I mentioned that in cosmology the only direct
signature from the time before chemical and thermal freeze-out 
seem to be the recently discovered small temperature fluctuations in
the microwave background. Generated by gravitation, 
%
%%%%%%%%%%%%%%%%%%%%%% Fig. 8 %%%%%%%%%%%%%%%%%%%%%%%%%%%%%%%%%%%%%%%%%%%%%%%
%\vspace*{-0.5cm}
\begin{figure}[ht]
\vspace*{7.2cm}
\includegraphics{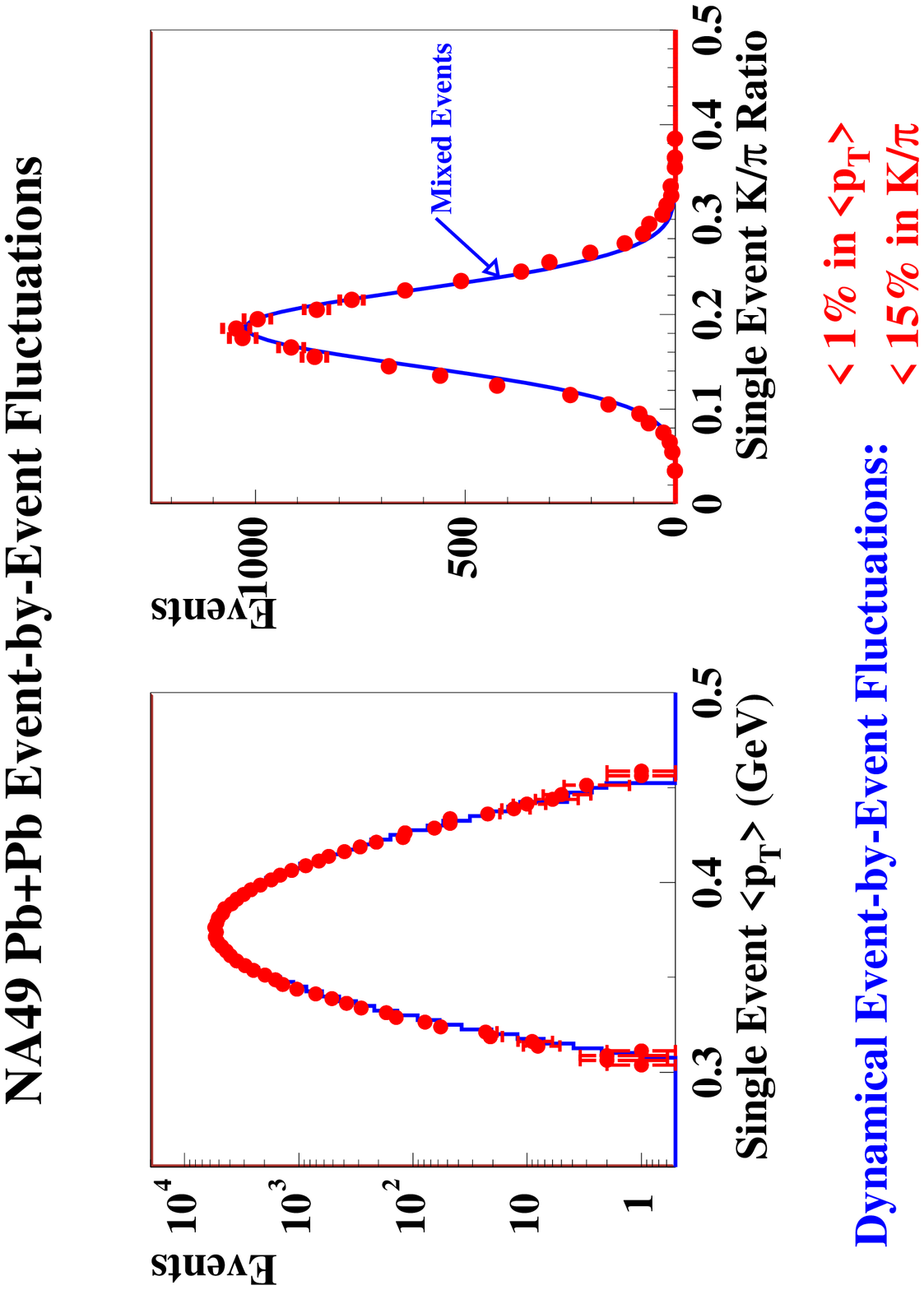}
%\centerline{\epsfxsize=10cm\epsffile{Fig8.eps}}
%\vspace*{-0.3cm}
\caption{Event-by-event fluctuations of $\langle p_\perp \rangle$ 
  and $K/\pi$ in Pb+Pb collisions at the SPS.\protect\cite{NA49QM97} 
  The solid lines denote the statistical fluctuations from mixed events.
\label{F8}}
\end{figure}
%\vspace*{-0.2truecm}
%%%%%%%%%%%%%%%%%%%%%%%%%%%%%%%%%%%%%%%%%%%%%%%%%%%%%%%%%%%%%%%%%%%%%%%%%%
%
they have no analogue in heavy-ion physics. There, however, we can 
study large numbers of collision events and investigate the 
fluctuations of physical observables from event to event. Such 
fluctuations are expected for statistical reasons because of the 
finite size of the collision system and the finite number of
produced particles. If the collision fireballs are truely thermalized,
finite number effects and quantum statistical fluctuations should be
the {\em only} reason for event-by-event fluctuations.\cite{Mrow}

Figure~\ref{F8} shows the measured distributions for the average
transverse momentum $\langle p_\perp\rangle$ and for the $K/\pi$ ratio
in individual Pb+Pb collision events at the SPS.\cite{NA49QM97}
These distributions are perfect Gaussians to a level of a few times
10$^{-4}$; moreover, the widths of these Gaussians agree very accurately
with expectations from mixed events, i.e. the fluctuations in the 
measured quantities from event to event are consistent with
purely statistical (e.g. thermal) fluctuations. No signs for large
``critical fluctuations'' expected near a phase transition or of other
dynamical fluctuations are visible. If, as argued above, the chemical
composition, which leads to a given average $K/\pi$ ratio, and the
thermalization and flow pattern, which causes a certain average
transverse momentum of the emitted hadrons, reflect interesting
physics of the fireball evolution, {\em every} Pb+Pb event at the SPS
reflects it in exactly the same way, at the level of less than a 
permille!  

%%%%%%%%%%%%%%%%%%%%%%%%%%%%%%%%%%%%%%%%%%%%%%%%%%%%%%%%%%%%%%%%%%%%%%%
\section{Conclusions}
\label{sec5}
%%%%%%%%%%%%%%%%%%%%%%%%%%%%%%%%%%%%%%%%%%%%%%%%%%%%%%%%%%%%%%%%%%%%%%%

I have presented arguments that in heavy-ion collisions at the SPS we 
have seen {\bf THE LITTLE BANG}: the measured soft hadron spectra show
convincing evidence for Hubble-like (3-dimensional) flow, 
for thermal (hadron) radiation, and for ``primordial'' 
hadrosynthesis out of a pre-hadronic state without quark confinement.
There are strong signs for the existence of a pre-hadronic stage
with non-trivial dynamics, resulting in a global strangeness 
enhancement by about a factor 2, in a particularly strong 
enhancement of multistrange (anti)baryons consistent with a
statistical hadronization picture, in the suppression of $J/\psi$,
$\chi_c$ and $\psi'$ states, and in elliptic flow. The 
absence of non-statistical fluctuations from event to event
indicates that all collision events are similar, that these
interesting phenomena occur in every event, and that decoupling
did not take place close to a phase-transition. For chemical decoupling
(which I argued to happen directly after hadron formation) this
implies that the hadronization process cannot be viewed as an
equilibrium phase transition; thermal freeze-out was found to occur
at much lower temperatures, in safe distance from any phase transition. 

The abundance and quality of hadron data now available allows for a
detailed and quantitative description of the final state in heavy-ion 
collisions. This begins to severely constrain dynamical extrapolations 
backward in time. Although no {\em direct} signatures from the 
pre-hadronic stage have so far been seen, the soft hadron data leave 
little room for doubt that it is there. Whether we can call it a 
quark-gluon plasma which satisfies the criteria spelled out in 
Sec.~\ref{sec1.2} will only become clear at higher collision energies 
where such direct signals become accessible. At the moment we only know 
that the pre-hadronic state
exerts pressure, so a certain degree of thermalization among the
partons must have occurred. The other important open question is, of
course: where is the threshold for deconfinement and how do we
identify it experimentally? Clearly, much work remains to be done but
now we know that we are on the right track.

%%%%%%%%%%%%%%%%%%%%%%%%%%%%%%%%%%%%%%%%%%%%%%%%%%%%%%%%%%%%%%%%%%%
\section*{Acknowledgments}
%%%%%%%%%%%%%%%%%%%%%%%%%%%%%%%%%%%%%%%%%%%%%%%%%%%%%%%%%%%%%%%%%%%
I would like to warmly thank the organizers of this stimulating workshop 
for the invitation and for their hospitality. I am grateful to C. Slotta 
and B. Tom\'a\v sik for help with the figures. This work was supported 
in part by GSI, DFG, and BMBF.

%%%%%%%%%%%%%%%%%%%% References %%%%%%%%%%%%%%%%%%%%%%%%%%%%%%%%%%

\end{document}